\documentclass{ws-mpla}
\usepackage{graphics,epsfig,wrapfig}
\newcommand{\be}{\begin{equation}}
\newcommand{\ee}{\end{equation}}
\newcommand{\ba}{\begin{eqnarray}}
\newcommand{\ea}{\end{eqnarray}}
\newcommand{\la}{\langle}
\newcommand{\ra}{\rangle}
\newcommand{\di}{ {\rm d} }
\renewcommand{\theequation}{\arabic{section}.\arabic{equation}}
\usepackage{color}
\definecolor{darkgreen}{rgb}{0,0.65,0}

\usepackage{amsfonts}
\usepackage{amssymb,latexsym}
\usepackage{amsmath,amsbsy}

\def\cdash{$^{\raisebox{-0.5pt}{\hbox{--}}}$}   

\begin{document}

\markboth{A.Efremov and A.Radyushkin}{Perturbative QCD of hard 
and soft processes}

%
\catchline{}{}{}{}{}
%

\title{ON PERTURBATIVE QCD OF HARD AND SOFT 
PROCESSES\footnote{Report of the Joint Institute for Nuclear 
Research E2-80-521, Dubna 1980 (unpublished). Submitted to the XX 
International Conference on High Energy Physics, Madison, July 
1980.}}

\author{A.V. EFREMOV$^\dag$ \and 
A.V. RADYUSHKIN$^{\dag\ddag}$}

\address{$^\dag$Joint Institute for Nuclear Research, Dubna, 
141980 Russia\\
$^{\ddag}$Present address: Old Dominion University,
Norfolk, VA 23529, USA\\ and \\ Thomas Jefferson National Accelerator Facility, 
Newport News, VA 23606, USA
}



\maketitle

\pub{Received (Day Month Year)}{Revised (Day Month Year)}

\begin{abstract}
We discuss some problems concerning the application of 
perturbative QCD to high energy processes. In particular for hard 
processes, we analyze higher order and higher twist corrections. 
It is argued that these effects are of great importance for 
understanding the behaviour of pion electromagnetic form factor 
at moderately large momentum transfers. For soft processes, we 
show that summing the contributions of the lowest twist operators 
leads to a Regge-like amplitude. 

\end{abstract}

\ccode{PACS numbers:  12.38.-t, 12.38.Cy, 12.39.St, 11.55.Jy }

\section{Introduction.}
\label{sec1} 

\begin{wrapfigure}[11]{R}{.6\textwidth}
\begin{center}
\vskip-7mm 
\includegraphics
[width=.55\textwidth]{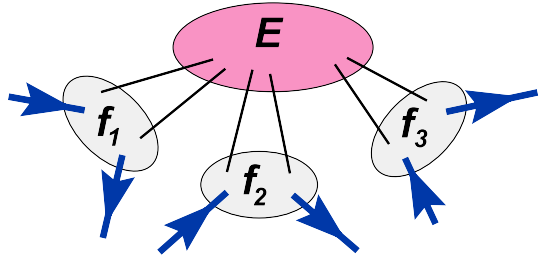}
\end{center}
\caption{\label{fig1}\footnotesize Illustration of Equation (\ref{eq1.1}).}
\end{wrapfigure}

It is now safe to assert that quantum chromodynamics (QCD) agrees 
qualitatively with all the experimental data related to strong 
interactions phenomena. At the same time, the QCD predictions are 
usually too flexible for precisive quantitative tests. This is 
caused mainly by the fact that all the calculations in QCD are 
based on perturbation theory (PT) , i.e. on the expansion over 
the coupling ``constant''  $\alpha_s(k)$ that depends really on 
the momentum scale $k$ related to the process investigated. 
Asymptotic freedom\cite{ref1} enables one to use PT at short 
distances (or large momenta). However, any physical process 
involves also long distances, i.e. each process involves small 
momentum scales $p^2$ (e.g., quark and hadron masses), and as a 
rule, this results in the appearance of the logarithmic 
contributions $\ln(Q^2/p^2)$, that are singular for $p^2=0$ (mass 
singularities\cite{ref2}\cdash\cite{ref4}). In such a situation 
$p^2$ cannot be neglected. However, within PT it is possible to 
show that for inclusive\cite{ref2}\cdash\cite{ref6} and some 
hadronic exclusive hard processes\cite{ref6}\cdash\cite{ref8} the 
$Q^2$-dependence of the corresponding amplitude $T(Q^2,p^2)$ can 
be factorized\footnote{({\it Comm. 2009.}) 
For the Yukava theories (scalar gluons) the property of 
factorization for hard processes was first considered even 
earlier 
(
A.~V.~Efremov,
Yad.\ Fiz.\  {\bf 19}, 196 (1974)).} 
from the $p^2$-dependence (see Fig.\ref{fig1}): 
\be
T(Q^2,p^2)= Q^N\left\{E(Q^2/\mu^2,\alpha_s(\mu))\otimes f(\mu^2,p^2)
+R(Q,p)\right\} \ , 
\label{eq1.1}
\ee 
where $N$ is the dimension of $T$ in mass units and $R$ is the 
sum of contributions which are power suppressed with respect to 
$E\otimes f$. The parameter $\mu$ is a boundary between large and 
small momenta, and $E\otimes f$ does not depend on a particular 
choice of $\mu$.  
 
The functions $f$ describe long-distance interactions. This means 
they cannot be reliably calculated in perturbative QCD and must 
be treated phenomenologically. As a result, the QCD predictions 
are more ambiguous. The functions $E$ describe short-distance 
interactions. In principle, they are given by a perturbative 
series expansion over $\alpha_s(\mu)$. In practice, only a few 
terms are known, and one is forced to make some plausible 
hypotheses about the magnitude of the non-calculated higher order 
corrections. According to most of recent estimates, 
$\alpha_s(\mu)/\pi$ is of the order 0.1 for $\mu^2\lesssim 
10\,{\rm GeV}^2$. This means that taking into account only a few 
first terms of the series is a good approximation only if the 
coefficients of the expansion of $E$ over $\alpha_s/\pi$ are not 
too large. 

An analogous uncertainty exists also for power corrections 
absorbed by $R(Q,p)$. It is known that the hadron and quark mass 
corrections for the most simple inclusive processes can be 
calculated exactly with the help of the $\xi$-scaling 
formalism\cite{ref9,ref10}. For light quarks $u\,,d\,,s$, their 
masses ($m_u\simeq4\,{\rm MeV},\, m_d\simeq7\,{\rm MeV},\, 
m_s\simeq120\, {\rm MeV}$) usually may be neglected at all. The 
main uncertainty is due to power corrections caused by the finite 
size of the hadrons, by the Fermi-motion of quarks inside the 
hadrons, etc. All these effects have a nonperturbative origin. 
The magnitude of the corresponding corrections $(M^2/Q^2)^n$ is 
determined by a characteristic scale $M\simeq 1/R_{\rm 
conf}\simeq300\div500\,{\rm MeV}$, but in some cases they play a 
very important role up to very high momentum transfers. For 
example, in high-$p_T$ hadron production, the effects of the 
primordial transverse momentum of partons dominate the cross 
section up to\cite{ref11} $p_T^2\simeq30\div 40\,{\rm GeV}^2$. It 
should be emphasized, however, that all the phenomenological 
methods of taking into account the power suppressed terms have no 
reliable theoretical basis. So, it is highly important to develop 
methods of field-theoretical analysis of power corrections. 

\section{Higher-order corrections} 
\label{sec2} 

\setcounter{equation}{0}

In general, the functions $ E(Q^2/\mu^2, \alpha_s(\mu))$ in Eq. 
(\ref{eq1.1}) depend on the calculation scheme, namely, on the 
chosen recipe of the R-operation for the ultraviolet divergences 
and the recipe of separating the contributions related to short 
and long distances (i. e. on the R-operation for composite 
operators). In particular, $ E$  is $\mu$-dependent. The 
functions $f(\mu^2,\, p^2)$ also depend on the chosen scheme and 
only the product $ E \otimes f$  is scheme-independent. If we 
take $\mu  = Q$, then the resultant expression would not have an 
explicit dependence on $\mu$. Note, that this procedure removes 
from $E(Q^2/\mu^2)$  the logarithms  $\ln(Q^2/\mu^2)$  which, for 
$Q \gg \mu$, lead to growth of the coefficients in the expansion 
of $E$ over $\alpha_s$. The meaning of the choice $\mu = Q$  is 
clear: one must  take $\mu$  equal to a scale characterizing the 
off-shellness of the particles taking part in the short-distance 
subprocess and the latter is proportional to $Q^2$:  $\langle k^2 
\rangle  \simeq -a^2Q ^2$. 

If, however, the parameter $a$ is very large (or very small) 
compared to 1, then the choice  $\mu \simeq  aQ$  should be 
preferred. It is implicit here that we use a ``physical'' 
renormalization scheme, i.e. $\bar g(k)$  corresponds to a vertex 
with external \mbox{momenta $k^2$.} However, for direct 
calculations within QCD it is more convenient to use various 
``unphysical''  schemes based on dimensional regularization. In 
this case the meaning of $\bar g(k)$  is less transparent. So, 
for the time being, we will adhere to the choice $\mu = Q$. 

Recently, in a series of papers\cite{ref12}\cdash\cite{ref15} it 
has been established that the scheme-dependence of the results 
obtained may be reduced by expanding the coupling constant 
$\alpha_s (Q)$  over $(\ln Q^2/\Lambda^2)^{-1}$,
\vskip-5mm
\begin{align}
 \frac{\alpha_s}{4\pi} =  \frac{4}{b_0\ln Q^2/ \Lambda^2}
\left \{1- \frac{b_1}{b_0^2} \frac{\ln \ln Q^2/ \Lambda^2} {\ln 
Q^2/ \Lambda^2 }  +\ldots \right\}\  ,\label{eq2.1} 
\end{align}
where $b_0,  b_1, \ldots $  are the coefficients of the expansion of the 
$\beta$-function over $g$.  After this change we have the following 
representation for $E \otimes f $:
\vskip-5mm
\begin{align}
 E \otimes f & = \left \{   (\ln Q^2/ \Lambda^2)^{-\gamma_0/b_0}
\sum_{k=0}^{\infty}\sum_{l=0}^{\infty} a_{lk}
 \frac{(\ln \ln Q^2/ \Lambda^2)^k}
{(b_0\ln Q^2/ \Lambda^2)^l }
  \right     \}\otimes \tilde f 
\nonumber \\
& =  \left \{   (\ln Q^2/ \Lambda^2)^{-\gamma_0/b_0}
 \left [ a_{00}   +   \frac{a_{10}}
{b_0\ln Q^2/ \Lambda^2 }
+   \frac{a_{20}}
{b_0^2 \ln^2   Q^2/ \Lambda^2 }
 \right.   \right.   \nonumber \\ 
 &  \hspace{4cm}\left.  \left. +   \frac{a_{21} \ln \ln Q^2/ \Lambda^2}
{b_0^2\ln^2 Q^2/ \Lambda^2 }  + \ldots 
 \right    ]
 \right     \}\otimes  \tilde f\ .\label{eq2.2} 
\end{align}

All information about the long-distance dynamics is accumulated 
in $ \tilde f $, whereas the coefficients  $a_{lk}$   can be 
calculated in PT. Moreover, the coefficients $a_{lk}^{(1)}$,  
$a_{lk}^{(2)}$ related to two different schemes can be obtained 
from one another by the change  $\Lambda_1 = \kappa_{12} 
\Lambda_2$  for the appropriately chosen parameter $\kappa_{12}$. 
Thus, if one uses the expansion (\ref{eq2.2}), then various 
schemes differ only in magnitude of the parameter $ \Lambda$. 

Let us assume that, in a scheme $S$, some first coefficients 
$a_{lk}$  are numbers of the order 1. Then in another scheme 
$S'$, which has  $ \Lambda' = 100\,\Lambda$  ( or $ \Lambda' = 
0.01\,  \Lambda$), the coefficients   $a_{lk}$ are numbers of the 
order $b_0 \ln 100 \simeq 40$. It is easy to notice the analogy 
with our previous discussion about the optimal choice of the 
parameter $\mu$ and to conclude that the scheme $S$  is very 
close to a ``physical'' scheme,  since the choice $\mu = Q $ 
(assumed in Eq. (\ref{eq2.2})) minimizes for this scheme the 
higher-order corrections. Note also that the choice $\mu=aQ$ is 
equivalent to an expansion over $(\ln(a^2Q^2/ \Lambda^2))^{-1}$ 
rather than over $(\ln Q^2/\Lambda^2)^{-1}$  , i.e. to the change 
$ \Lambda \to  \Lambda/a$. 

So, let us assume that if one uses a physical scheme with a 
properly chosen subtraction point (i.e., $\mu^2 = \langle 
k^2\rangle$), then the resultant expansion over $ 
(\alpha_s/\pi)$  has coefficients of the order 1 (this is just 
the situation usually encountered in QED, where one has no 
problems with the momentum-dependence of the coupling constant 
$\alpha_{\mbox{\tiny QED}} \simeq 1/137$). If this assumption is 
valid, then the higher-order corrections may be calculated using 
the following rules: 

1) One may calculate in an arbitrary scheme. The most 
convenient,  in our view, is the $\overline{\rm 
MS}$-scheme\cite{ref16}, which is free  from spurious terms $\ln 
(4\pi) $ and $\gamma_E$  present in the minimal-subtraction  
(MS)-scheme. The parameter $\Lambda_{\overline{\rm MS}}$ may be 
chosen to be a fundamental scale of QCD. Note, that   
$\Lambda_{\overline{\rm MS}}$ is close  to  
$\Lambda_{\overline{\rm PH}}$ related to a physical scheme: 
$\Lambda_{\overline{\rm PH}}= \kappa\Lambda_{\overline{\rm MS}}$, 
where $\kappa \simeq  2$  is almost independent of the vertex 
chosen to define $\bar g(k)$ (cf. Ref. \refcite{ref15}). 
   
2) In general, however, there are no a priori reasons to expect 
that the \mbox{ $\overline{\rm MS}$-scheme}  minimizes the 
coefficients $a_{lk}\otimes f$ in Eq. (\ref{eq2.2}). If it is 
known that the average off-shellness of lines related to a 
short-distance subprocess is $a^2Q^2$ and $a\gg 1$ ( or $a\ll  
1$),  then $E\otimes  f$  must be expanded over $\left 
[\ln(a^2Q^2/\kappa^2 \Lambda^2_{\overline{\rm MS}}) \right 
]^{-1}$ rather than over $\left [\ln (Q^2/ 
\Lambda^2_{\overline{\rm MS}})\right ]^{-1}$

3) Usually the value of $a$ is not known. However, this value may 
be estimated by requiring that the coefficient $a_{10}$  (or 
$a_{20}$, if $\gamma_0=0$) vanish after the change $ 
\Lambda_{\overline{\rm MS}} \to 2\Lambda_{\overline{\rm MS}}/a $. 

In this approach all results of the calculations are expressed in 
terms of the only parameter $\Lambda_0=2\Lambda_{\overline{\rm 
MS}}$. However in the expansion (\ref{eq2.2}) for different 
processes one may use different $\Lambda_{\rm eff}^{(i)} = 
\Lambda_0 /a_i$ (with known $a_i$'s). 

\section{Power corrections} 
\label{sec3} 

\setcounter{equation}{0}

Our derivation of Eq. (\ref{eq1.1})  given in 
Refs.~\refcite{ref5}--\refcite{ref7} is based on the analysis of 
Feynman diagrams in the $\alpha$-representation\cite{ref17} (see 
also Refs.~\refcite{ref18,ref19}), i.e. on the formula 
\begin{align}
\frac1{m_\sigma^2 - k^2} = i \int_0^\infty d\alpha_\sigma \, e^{i 
\alpha_\sigma (k^2-m_\sigma^2)} \label{eq3.1}
\end{align}
applied to propagators of all lines $\sigma$ of the diagram. After 
integrations over all $k_i$ this gives the representation 
\begin{align}
T(Q,p)\sim &\int\limits_0^\infty \prod\limits_\sigma 
\di\alpha_\sigma\,D^{-2}(\alpha)\, G(Q,p,\alpha)\nonumber \\ 
&\times \exp\left[ iQ^2 A(\alpha)+i p^2 I(\alpha) -
i\sum_\sigma \alpha_\sigma m_\sigma^2 \right]\,,\label{eq3.2} 
\end{align}
which has many advantages for analysis of the large $Q^2$ 
behaviour of $T$. In particular, from Eq. (\ref{eq3.2}) it 
follows that integration over a region where $A(\alpha) > \rho$  
gives for $Q^2\to \infty$ an exponentially damped contribution 
${\cal O}[\exp (-Q^2 \rho)]$. Hence all contributions having a 
power (${\cal O}(Q^{-N})$) behaviour for $Q^2 \to \infty $ are 
due to integration over regions where $A(\alpha)$ vanishes. There 
exist three main possibilities to get   $A(\alpha) = 0$: 

1) short-distance (or small-$\alpha$)  regime, when  
$\alpha_{\sigma_1}= \alpha_{\sigma_2} = \ldots =  
\alpha_{\sigma_n} = 0$ for some lines $\sigma_1, \sigma_2, \ldots 
, \sigma_n$;

2) infrared (or $\alpha \to \infty$) regime, when 
$\alpha_{\sigma_1}= \alpha_{\sigma_2} = \ldots =  
\alpha_{\sigma_n} = \infty$ for a set of lines $\{ \sigma_1, 
\sigma_2, \ldots , \sigma_n \}$; 

3) pinch regime, when $A(\alpha)=0$ for nonzero finite 
$\alpha$'s. This regime works when $A(\alpha)$ may be represented 
as a difference of two positive terms. 

It is possible also to get  $A(\alpha) = 0$  making up a 
combination of the three basic regimes. In the momentum 
representation,  the first regime corresponds to integration over 
a region $k \sim Q$, the second one over $k \sim p^2/Q$, and the 
third over $k \sim p$. This  means that perturbative QCD is 
applicable only when the regimes 2,3 and the combined regimes 
either do not contribute at all or give a power suppressed 
contribution compared to that of the pure SD-regime.There exists 
a wide class of processes for which the pinch regime does not 
work (see Ref.~\refcite{ref5}), and it is sufficient to analyze 
only the SD- and IR-regimes. In this case it is very useful to 
visualize a diagram as an electric circuit and to treat the 
parameters $\alpha_\sigma$ as the resistances of the 
corresponding lines $\sigma$. Note, that according to Eq. 
(\ref{eq3.2}), for $A(\alpha) = 0$ the amplitude $T$ lacks its 
$Q^2$-dependence. Hence, one must find the subgraphs that should 
possess the following topological properties: when lines of these 
subgraphs are contracted into point ($\alpha_\sigma= 0$)  and/or 
removed from the diagram ($\alpha_\sigma= \infty$) then the 
resulting diagram does not depend on $Q^2$. Each configuration of 
this type corresponds to some power-behaved ${\cal 
O}(Q^{-N})$-contribution. The power $N$  may be easily estimated 
with the help of the rules $k_{\rm SD} \sim  Q$, $k_{\rm IR} 
\sim   p^2/Q$: 
\begin{align}
& t_V^{\rm SD}  \lesssim Q^{4 -\sum t_i} \  ; \nonumber \\ 
& t_S^{\rm IR}  \lesssim Q^{ -\sum t_j} \  ;\label{eq3.3}\\ 
& t_{V;S}^{\rm SD;IR}\lesssim Q^{4 -\sum t_i-\sum t_j}\ ,\nonumber 
\end{align}
where $t_i$  ($t_j$) is twist\cite{ref20} of the $i$-th ($j$-th) 
external line of the subgraph $V$ ($S$)  corresponding to the 
SD-(IR-) integration. Recall that $t_{i,j} =1$  for $\psi, \bar 
\psi$-fields and the curl $G_{\mu \nu}$, whereas  $t_{i,j} =0$  
for the vector field $A_\mu$. That is why in QCD (in covariant 
gauges) it is necessary to sum up over external gluon lines of 
the subgraphs $V, S$.
 
However, for the forward amplitudes (corresponding to inclusive 
cross sections) and for amplitudes of exclusive processes 
involving colour singlet particles, after such a summation the 
field $A_\mu$ either disappears (and the gluon lines correspond 
to the curl $G_{\mu\nu}$ that has a nonzero twist) or enters into 
covariant derivatives \mbox{$D_\mu = \partial_\mu - igA_\mu$}  
present in composite operators that naturally arise when the 
contribution of the corresponding configuration is written in the 
coordinate representation (cf. 
\mbox{Refs.~\refcite{ref5}--\refcite{ref7}).} 

Consider, e.g. the forward amplitude $T(\tau,Q^2)$  corresponding 
to the total cross section of the Drell-Yan process $AB \to \mu^+ 
\mu^- X$. In this case all the configurations responsible for the 
leading contribution $T^{\rm lead}  (\tau, Q^2) = {\cal O} 
((Q^2)^0)$ have the structure shown in Fig. \ref{fig2}a. Here, 
the subgraph $V$ corresponds to the $E$-function, whereas 
subgraphs resulting after contraction of $V$ into point 
correspond to the function \mbox{$f = f_A \otimes f_B$.}  The 
configurations shown in Fig. \ref{fig2}b,c give power suppressed 
contributions. Note also that Eq. (\ref{eq3.3}) gives only an 
upper estimate. This means the contribution Fig. \ref{fig2}a 
itself apart from the leading contribution, contains also power 
corrections. These corrections appear in the following cases: 
\begin{figure}[t]
\begin{center}
\includegraphics[width=.7\textwidth]
{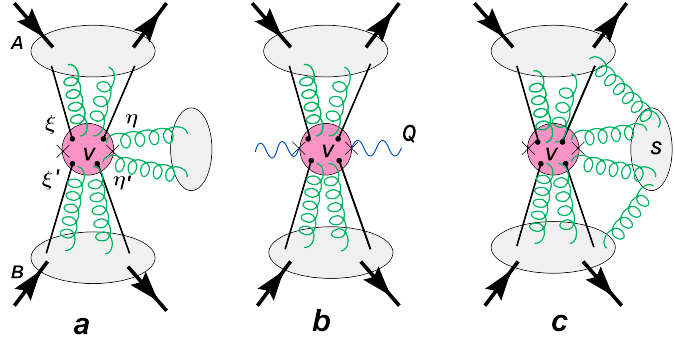}
\end{center}
\caption{\label{fig2}\footnotesize 
The configurations for leading contributions to Drell-Yan process 
with large $Q^2$ .}
\end{figure}

1) If we project the spinor structure of the subgraph $V $ onto 1 
rather than onto $\gamma_\mu$. In the latter case we obtain a  
composite operator having twist equal to 2 and in the former  one 
the resulting operator has twist 3. Note that $\gamma_5, 
\gamma_5\gamma_\mu$ and $\sigma_{\mu \nu}$-projections have 
vanishing matrix elements for spin-averaged amplitudes. 

2) If we expand a bilocal operator ${\cal O}^\mu (\xi, \eta)$ 
over the local ones, then there appear operators 
\begin{align}
& \ \ \ \ j \ {\rm times} \nonumber \\
 \biggl \{ \bar \psi  \left ( \gamma^\mu D^{\mu_1} \ldots D^{\mu_n}
\right )_{\rm symmetrized} & {\overbrace{g_{\mu . \mu .} \ldots 
g_{\mu . \mu .} }} \, \psi \biggr \}\label{eq3.4}
\end{align}
having twist $2 + 2j$. Each factor $g_{\mu . \mu .}$ adds $(\xi - 
\eta)^2$ to the corresponding function $E(\xi,\eta;  \xi', 
\eta')$ and this leads to suppression of the resulting 
contribution by an additional factor $1/Q^2$.  
    
The expansion of the function $E(\xi,\eta, \ldots)
(\xi-\eta)^{\mu_1} \ldots (\xi-\eta)^{\mu_n}$ over 
symmetric-traceless structures $E(\xi,\eta, \ldots) \left 
[(\xi-\eta)^2 \right ]^j \left \{(\xi-\eta)^{\mu_1} \ldots 
(\xi-\eta)^{\mu_l} \right \}$ corresponds in the momentum 
representation to an expansion of the amplitude $E(k, Q,\ldots)$ 
\mbox{over $k^2$: } 
\begin{align}
\tilde E (k, \ldots)|_{k^2=0} + \sum_{j=1}  (k^2)^j  \tilde E_j 
(k, \ldots)|_{k^2=0} \label{eq3.5} \  , 
\end{align}
i.e. over the off-shellness of the particle corresponding to the 
external line of the SD-subgraph $V$ (quark masses are assumed to 
be zero). In the $\alpha$-representation this corresponds to an 
expansion of the integrand in Eq. (\ref{eq3.2}) into a power 
series over  $\lambda_V=\sum\limits_{\sigma\in V}\alpha_\sigma$. 
Thus, the coefficient function $E$ for the leading contribution 
corresponds to an on-shell amplitude. This means that $E$ is 
formally gauge-invariant in each order of PT. However, because of 
logarithms $\ln Q^2/k^2$  present in $\tilde E (k, Q •.•. ) 
$, taking the limit   $k^2=0$  is a rather delicate procedure. 
Note, that $E(Q^2/\mu^2, \ldots )$ in Eq. (\ref{eq1.1}) 
corresponds to integration over small  $\lambda_V$, i.e. the 
contribution of the region $\lambda_V>1/\mu^2$ must be subtracted 
off. As a result, one has $\ln Q^2/\mu^2$ in place of $\ln 
Q^2/k^2$, and it is then safe to take $k^2=0$.• To maintain 
gauge invariance, one may introduce the cut-off at 
$\lambda_V>1/\mu^2$  using, e.g., the dimensional regularization 
$d^4 k \to d^{4+ 2\epsilon }k (\mu^2)^{-\epsilon}$  combined with 
subtraction of poles $1/\epsilon$. These poles formally 
correspond to $\ln (\mu^2/k^2)|_{k^2=0} $ (cf. 
Ref.~\refcite{ref23}). 

Thus in the leading power approximation quarks corresponding to 
external lines of the parton subprocess should be treated as the 
on-shell particles. Real quarks are, of course, off-shell. But 
according to \mbox{Eqs.~(\ref{eq3.4}), (\ref{eq3.5}) } this 
phenomenon leads to power corrections only. They may be analyzed 
just in the same way as the leading term, although the analysis 
is more involved. In particular, for each new set of operators 
one must introduce a new function. However some of these 
functions are linearly dependent due to equations of motion. As 
it was emphasized in a classic paper\cite{ref24}, using equation 
of motion \mbox{$D_\mu \gamma^\mu\psi = 0$}, one may get rid of 
the operators containing $D_\mu \gamma^\mu$ and $D_\mu D^\mu$.  
The resulting operators are built of the fields $\psi, \bar \psi$ 
and covariant derivatives $D_\mu$. The reduced matrix elements of 
such operators may be identified with the moments of functions 
which are generalizations of the parton distribution functions, 
e.g. 
\begin{align}
&
\langle P| \bar \psi_a 
\left \{ \left ( \sigma_{\mu \nu}
G^{\mu \nu}_{; \mu_1 \ldots \mu_k} \right )
\psi_{; \mu_{k+1} \ldots \mu_{n+k}}
\right \} |P \rangle 
 = \frac{1+(-1)^k}{2} \left [ \tilde f_{a,g} (n,k) +(-1)^n  \tilde 
f_{\bar a,g} (n,k) \right ]  \nonumber \\ 
&  \hspace{3mm} = \frac12  \left [ 
\tilde f_{ag,a} (n,k) +(-1)^k  \tilde f_{a,ga} (n,k) +(-1)^n  
\tilde f_{\bar ag,\bar a} (n,k) +(-1)^{n+k}  \tilde f_{\bar a, g 
\bar a} (n,k) \right ] \nonumber \\ 
& \hspace{3mm}=\!\!\int\limits_0^1\!\di x_1\di x_2\di x_3\!\biggl[\!f_{ag,a} 
(x_1 ,x_2; x_3) \delta 
(x_1\!\!+\!\!x_2\!\!-\!\!x_3)\!+\!(\!-1)^k\!
f_{a,ga}(x_1;x_2,x_3)\delta(x_1\!\!-\!\!x_2\!\!-\!\!x_3) 
\nonumber \\ &  
\hspace{3cm} +(-1)^n  \ldots +(-1)^{n+k}  \ldots \biggr ] 
x_1^{n-1} x_2^{k-1} \  ,\label{eq3.6}
\end{align}
where ;  denotes covariant differentiation, $a$ denotes the quark 
flavour and $\bar a$ that of antiquark. The function  $f_{ag,a}$ 
($f_{\bar a,g\bar a}$) describes a quark (antiquark) with 
momentum $x_1P$ and a gluon with momentum $x_2P $ in the initial 
state and quark (antiquark) with momentum $x_3P=(x_1+x_2)P$ in 
the final state. The functions $ f_{ag,a},f_{\bar ag,\bar a}$ 
have analogous meaning. Such a construction was introduced first 
in Ref.~\refcite{ref25} for operators $\bar \psi \left \{  
\gamma_{\mu}\left (\partial_{\mu_1} \ldots \partial_{\mu_k} 
\right ) \left (\partial_{\mu_{k+1}} \ldots \partial_{\mu_{n+k}} 
A_{\mu_{n+k+1}}\right ) \right \} \psi $ used in the analysis of 
factorization in the Feynman gauge.

Operators of Eq. (\ref{eq3.6}) appear also in configurations 2a 
if the subgraph $V$ has external lines corresponding to the curl 
$G_{\mu\nu}$. Apart from matrix elements $\langle P |{\cal O}  
|P\rangle$ these configurations  contain also matrix elements  
$\langle 0 |G \ldots G|0\rangle$, $\langle 0|(\bar \psi \psi) 
\ldots (\bar \psi \psi )  |0\rangle$ $\langle 0 |G \ldots G (\bar 
\psi \psi )|0\rangle$, etc. In each order of PT these matrix 
elements vanish, but in QCD, due to nonperturbative effects, 
these vacuum matrix elements may be nonzero. As it was 
demonstrated in Ref.~\refcite{ref26}, these contributions are 
very important for understanding the dynamics of hadrons. The 
main problem is to generalize the methods developed in 
Ref.~\refcite{ref26} to more complicated amplitudes. 

All the configurations considered above correspond to the 
SD-regime $\lambda (V) \sim 0$.  One must take into account also 
the configurations 2c corresponding to the combined SD-IR regime 
$\lambda (V) \sim 0$, $\lambda (S) \to \infty $. Physically this 
regime corresponds to a short-distance subprocess accompanied by 
the exchange of soft quanta between the hadrons $A$ and $B$. 
According to Eq. (\ref{eq3.3}), these contributions also have a 
power behaviour ${\cal O} (Q^{-N})$, where $N$ is the number of 
external lines of subgraph $S$. If all $N$ lines are gluonic, 
then the corresponding diagrams describe a multipole interaction 
of hadrons. However, if the quarks are massless, then the 
subgraph may possess quark lines also. The main contribution for 
the IR-regime is given by the region $k^2 \sim  (p^2/Q)^2$, where 
$p^2$ may be treated as hadronic mass. For massive fields (e.g., 
quarks) the contribution of the IR-regime is damped by the mass 
present in the propagator $(k^2+m^2)^{-1}$ if $k^2 \sim m_H^4/Q^2 
\ll  m^2$, i.e. for $Q^2 \gtrsim m_H^4/m^2$, the IR-regime does 
not work. The mass $m$ in this case works as an infrared cut-off. 

Since the gluons are massless, the IR-regime in PT always works 
for gluons, and there are power corrections due to integration 
over $\alpha \to \infty$. However, the contributions 
corresponding to the configuration 2c do not factorize in the 
usual sense. This suggests that complete analysis even of the 
lowest power corrections in perturbative QCD is impossible. 
However, if the exchanged system is coloured, then the 
corresponding contribution should be damped by confinement 
(nonperturbative) effects, i.e. in this case even for $m=0$ there 
exists an IR cut-off $M \equiv 1/R_{\rm conf} \sim  300 \div 
500$\, {\rm MeV}. Thus, if we add to PT a confinement hypothesis, 
then for a coloured system $S$ the contribution of configuration 
2c is damped for $Q^2 \gtrsim  m_H^4/M^2$, i.e. for all hard 
processes. On the other hand, if the exchanged system is 
colour-singlet (e.g., colour-singlet glue-ball, $\pi$-meson, 
$\rho$-meson,  pomeron etc.), then there are no \mbox{\it a 
priori} grounds to neglect the configuration 2c. We feel that the 
methods of the ``old'' hadronic theory, such as the Reggeon 
calculus and potential models (e.g. the quasi-potential 
approach\cite{ref27})  will be much more suitable for analysis of 
these contributions than the perturbative QCD methods. Highly 
instructive in this connection is the result of 
Ref.~\refcite{ref28}, where it is shown that if one describes the 
soft exchange by an exponentially vanishing quasi-potential, then 
the soft interactions in initial and final states give power 
($1/p_T$  and $1/p_T^2$)  rather than exponential ($\exp(-ap_T 
)$) corrections to the amplitude of high-energy wide-angle 
elastic $\pi p$-  and $pp$-scattering. What is more, in the 
available energy range these corrections give an essential 
contribution. 

Thus, a consistent analysis of power corrections in QCD 
seems to be a highly nontrivial but maybe not a hopeless task. 

After this paper was essentially completed, we received a 
preprint by Politzer,\cite{ref29} where the power corrections are 
analyzed using the methods similar to ours. There is no surprise 
that the analyzes are similar, because both are based on the 
classic work.\cite{ref24} However, there exists also a conceptual 
difference between the two approaches. Our approach is based 
directly on the analysis of the corresponding amplitude 
$T(Q^2,p^2)$ in the large-$Q^2$ region whereas Politzer's 
approach is based on the analysis of small-$p^2$ behaviour of  
$T(Q^2,p^2)$ (i.e. on the analysis of mass singularities). Both 
the approaches are (almost) equivalent if one analyzes 
logarithmic $\ln (Q^2/p^2)$-corrections. However, the power 
corrections ${\cal O}(Q^{-2N})$ in mass-singularity analysis 
correspond to contributions $(p^2)^N (\ln p^2)^k$, that simply 
vanish for $p^2= 0$. So, we are very sceptic about the main idea 
of Ref.~\refcite{ref29} that a complete analysis of power 
corrections may be performed within the mass-singularity 
approach. Incompleteness of this approach reveals itself in the 
fact that the soft exchanges (configuration 2c) are completely 
ignored in  Ref.~\refcite{ref29}. Another disadvantage of the 
mass-singularity approach is that it incorporates perturbation 
theory just in the region $k^2\sim0$,  where one should expect in 
QCD large nonperturbative effects. In particular, within the 
mass-singularity approach, it seems impossible to understand the 
origin and structure of power corrections to the ``$e^+e^- \to 
{\rm hadrons}$'' process, since the related amplitude $T(Q^2)$ 
does not depend (in a massless theory) on small momentum 
variables like $p^2$. It should be emphasized that the analysis 
of these effects given in Ref.~\refcite{ref26} is the only 
serious analysis of power corrections in QCD and it is based (not 
by chance) on the operator product expansion, i.e. just on the 
analysis of the large-$Q^2$ behaviour of the relevant amplitude. 

\section{Pion form factor at moderately  large $Q^2$}
\label{sec4} 
 \setcounter{equation}{0}

As an example illustrating the importance of a detailed study of 
higher order and higher twist effects, let us consider the 
behaviour of pion electromagnetic form factor $F_\pi (Q) $ for 
moderately large $Q^2$ 

\begin{wrapfigure}[10]{R}{.5\textwidth}
\begin{center}
\vskip-7mm 
\includegraphics
[width=.45\textwidth]{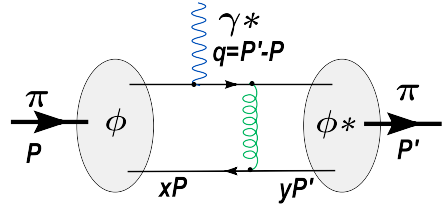}
\end{center}
\caption{\label{fig3}\footnotesize Diagram for asymptotic term of pion
form factor.}
\end{wrapfigure}

During the last 3 years a definite progress has been made in 
understanding of the asymptotical $Q^2 \to \infty$ behaviour of  
$F_\pi (Q)$ in the QCD 
framework.\cite{ref6}\cdash\cite{ref8,ref30}\cdash\cite{ref33}
The main result here is the proof that in a region where the 
power corrections may be neglected the form factor may be written 
in a factorized form\cite{ref6}\cdash\cite{ref8}  
\begin{align}
F_\pi (Q) =& \frac1{Q^2} \int\limits_0^1 \di x \int\limits_0^1 
\di y \, \varphi^* (y, \mu^2,\mu_R^2, \alpha_s (\mu_R)) \nonumber 
\\ & \times E(Q^2/\mu^2, \mu_R^2/\mu^2,x,y,  \alpha_s (\mu_R))  
\, \varphi (x, \mu^2,\mu_R^2, \alpha_s (\mu_R))\,,\label{eq4.1}
\end{align}
where $\varphi (x)$  is the wave function describing the 
splitting of the pion into a $q\bar q$-state and $E/Q^2$  is the 
amplitude of the short-distance subprocess $\gamma^* q\bar q \to 
q' \bar q'$. Note, that, in general, the normalization parameter 
$\mu_R$ of the R-operation for ordinary UV -divergences may 
differ from the splitting parameter $\mu$  that separates small 
and large momenta. The latter may be treated also as the 
normalization parameter for composite operators. The moments of 
the function   $\varphi (x)$ are equal to the reduced matrix 
elements of the twist-2 operators $\bar\psi \gamma_\mu D^n \psi$. 
The SD-amplitude, as usual, is a series expansion over $\alpha_s 
(\mu_R)$  (see Fig. \ref{fig3}): 
 \begin{align}
E(Q^2/\mu^2, \mu_R^2/\mu^2,x,y,  \alpha_s (\mu_R))  =
 \frac{2 \pi  \alpha_s (\mu_R) C_F}{N_c} \cdot
\frac{xQ^2}{(xQ^2) (xyQ^2)} \left \{ 1 +{\cal O}(\alpha_s) \right 
\} \,,\label{eq4.2}
\end{align}
where $C_F = 4/3$,  $N_c = 3$.  The factor $xQ^2$ in 
the numerator of of Eq. (\ref{eq4.2}) is due to the trace over 
Dirac $\gamma$-matrices. Note that $E (x, y)$ is rather singular 
for $x, y\to 0$. Hence, the main contribution is given by 
integration over small $x,y$. In the next order the most singular 
terms are $\frac1{xy}  \ln (xyQ^2/\mu_R^2)$ and $\frac1{xy}(\ln x 
) (\ln y) $,\ $(\ln Q^2/\mu^2) \cdot (\ln xy)/(xy)$. The first  
term is given by divergent parts (notice $\mu_R$) and the others 
by convergent ones. Thus, to minimize the $\alpha_s$-corrections 
we must take  $\mu^2 \simeq \bar x Q^2 = \bar y Q^2$  and 
$\mu_R^2 \simeq \bar x \bar y Q^2$, where $\bar x$  ($\bar y$) is 
the average value of $x$ (or $y$): 
\begin{align}
\label{eq4.3}
\ln \bar x = \langle \ln x \rangle \equiv 
\left(\int\limits_0^1\ln x\,\frac{\varphi(x)}{x}\,\di x\right) 
\left(\int\limits_0^1\frac{\varphi(x)}{x}\,\di x\right)^{-1} 
\end{align}
If $\varphi(x)\sim \delta\left(x-\frac12\right)$ (noninteracting 
quarks), then $\bar x=\frac12$. However, for very broad 
functions, e.g. for $\varphi(x)\sim[(x(1-x)]^R$ with $R\ll 1$,  
we have a very small value $\bar x\sim\exp(-1/R)$.  

In the problem investigated we encounter just the same mass 
scales ($m_\pi, m_q$  and $M=1/R_{\rm conf}$) as in deep 
inelastic scattering. So, it seems natural to expect that Eqs. 
(\ref{eq4.1}), (\ref{eq4.2}), for the appropriately chosen wave 
function $\varphi(x)$ must provide a good approximation for 
$F_\pi(Q)$ in the region $Q\gtrsim 1$\,{\rm GeV}. The wave 
functions  $\varphi (x,\mu^2)$ are in general unknown. In 
perturbative QCD one may calculate only their evolution with 
growing $\mu^2$. In particular $\varphi(x,\mu^2)\to 6f_\pi x(1 
-x)$ as $\mu^2\to \infty$, where $f_\pi=133\, {\rm MeV}$. 
Presence of the $f_\pi$-factor is due to the normalization 
condition 
\begin{align}
  \int\limits_0^1\varphi(x,\mu^2)\,\di x= f_\pi\,.
  \label{eq4.4}
\end{align}

It is clear, however, that for $\mu\lesssim 1\,{\rm GeV}$ the 
wave function $\varphi(x,\mu^2)$  may strongly differ from its 
limiting form. For noninteracting particles $\varphi (x) \sim 
\delta (x -1/2)$. When the interactions are switched on, the wave 
function broadens. The width $\Gamma$ of $\varphi (x) $ may be 
estimated as $\Gamma \sim (E_{\rm int}/m_q)^2$. Hence, for heavy 
mesons (e.g. for  $J/\psi$  or $\varUpsilon$-particles) $\varphi 
(x)$ is rather narrow, since $E_{\rm int}\simeq M = 300\div 
500\,{\rm MeV}$ and $m_q > 1\,{\rm GeV}$. On the other hand, for 
pions the wave function must be very broad, because\cite{ref26}  
$m_u \simeq 4\,{\rm MeV}$, $m_d\simeq 7\,{\rm MeV}$, i.e. pion 
must be treated as an ultra-relativistic system. To obtain a more 
accurate estimate of the width $\Gamma$ for such a system we 
assume that the (soft) Bethe-Salpeter wave function 
$\chi_P(k_1,k_2)$ is exponentially damped for moderately large  
spacelike $k_i^2$ ($i=l,2$): 
\begin{align}
 \chi_P(k_1,k_2) \sim \frac1{k_1^2 k_2^2}\,\exp (k_i^2/M^2)\,; 
\ \ -k_i^2 \gtrsim M^2\,.\label{eq4.5}
\end{align}
The exponential damping is suggested by the observed spectra of 
particles produced in high-energy hadronic reactions. For our 
wave function $\varphi (x)$ (which may be obtained from 
$\chi_P(k_1,k_2)$ by integration\cite{ref35,ref36} over $k_-=k_0 
-k_3$ and $k_{\perp}$),  the choice (\ref{eq4.5}) gives 
\begin{eqnarray}
  \varphi (x,\mu^2\sim M^2)  \simeq f_\pi \left \{
\begin{array}{lll} \exp[-m_u^2/xM^2]  \ ; & x  \ll1  \\\label{eq4.6}  \\
  \exp[-m_d^2/(1-x)M^2]  \ ;& (1-x) \ll 1 \  .
 \end{array}
  \right.
\end{eqnarray}
Thus, $ \varphi (x)$ is very close to $f_\pi$  everywhere outside 
the regions \mbox{$0 \leq  x \leq  m_u^2 /M^2$} $  \sim 10^{-4}$;    
$0 \leq 1-x \leq m_d^2/M^2 \sim 10^{-3}$. In these regions $ 
\varphi (x)$ vanishes rapidly. Note that for such a wave function 
$\bar x\sim  10^{-3} \div 10^{-4}$, i.e. the main contribution 
into $F_\pi$  is given by the region where the gluon has a 
catastrophically small off-shellness  $\bar x^2  Q^2$, which for 
$Q^2 < 100$\,{\rm GeV}$^2$ is much smaller than the value $|k^2| 
\sim 0.1 \div  0.3$\,{\rm GeV}$^2$, where the confinement effects 
must be taken into account. Thus, for a broad wave function short 
distances do not contribute, in fact, and Eq. (\ref{eq4.2}) is 
unreliable. In particular, it is not justifiable to neglect the 
power corrections that may really have a form $(M^ 2/ \langle k^2 
\rangle) \sim (M^2/\bar x Q^2)$ rather than simply $M^2/Q^2$. We 
assume that the confinement effects eventually remove the 
infrared singularity from the ``hard''  quark and gluon 
propagators $1/xQ^2$ and $1/xyQ^2$.  So, we change \mbox{$1/xyQ^2 
\to 1/(xyQ^2+2M^2)$} $ \sim1/(xyQ^2+ \langle (k_\perp - 
k_\perp')^2 \rangle )$  and $1/xQ^2 \to  1/(xQ^2+M^2)  \sim 
1/(xQ^2+ \langle k_\perp^2 \rangle )$. The connection between 
$M^2$ and $\langle k_\perp^2 \rangle$ is a pure mnemonics and 
must not be understood too literally. However, as an 
order-of-magnitude estimate this connection must be true. So, we 
should expect that $M^2\simeq 0.1 \div 0.3\,{\rm GeV}^2$. 

As a result, we have in place of Eqs. (\ref{eq4.1}), 
(\ref{eq4.2}) 
\begin{align}
 F_\pi^{AA} (Q) =  \frac{2\pi C_F}{N_c} \int\limits_0^1 \di x\di y \, 
\varphi(x)\,\varphi(y)  &  \frac{ \alpha_s (\mu_R^2) \, xQ^2}
{(xQ^2 +M^2)(xy Q^2 +2 M^2)}  \nonumber \\
&\times\left\{1+{\cal O} (\alpha_s)\right \}\,,\label{eq4.7}
\end{align}
where $AA$ stands for projection onto the ``axial'' operators 
$\bar\psi\gamma_5\gamma_\mu D^n\psi$.

From  Eq. (\ref{eq4.7}) it is clear that for not too large $Q^2$  
the contribution of the soft region $x\sim  0$ is damped at 
\mbox{$x \sim M^2/Q^2$,} whereas the wave function damps only the 
region $x \lesssim m_q^2/M^2$. Thus, up to $Q^2 \sim(M^2/m_q)^2 
\gtrsim 10^3$\,{\rm GeV}$^2$ the magnitude of the pion form 
factor is determined by  the value of $M^2$, i. e. by the 
confinement radius. 

The main contribution into the integral in Eq. (\ref{eq4.7}) is 
given by the region \mbox{$xyQ^2 \sim  2M^2$.}  To minimize the 
$\alpha_s$-correction we should take $\mu_R^2$ equal to the 
average off-shellness of the gluon: $\mu_R^2 =\bar x \bar y 
Q^2+2M^2 \simeq 4M^2$,
\begin{align}
 \alpha_s(\mu_R^2)\to\alpha_s(4M^2)=\frac{4\pi}{9\ln(4M^2/
\Lambda^2_{\rm PH})}\,.\label{eq4.8}
\end{align}

It should be realized that Eqs. (\ref{eq4.7}), (\ref{eq4.8}) are 
meaningful only if $\alpha_s(4M^2)/\pi\ll 1$, i.e. for 
$4M^2/\Lambda^2\gtrsim 50$ (in this case $\alpha_s(4 M^2)/\pi 
\simeq 0.1$). If $M^2\sim 0.2$\,{\rm GeV}$^2$, then Eqs. 
(\ref{eq4.7}), (\ref{eq4.8}) may work only for $\Lambda_{\rm 
PH}\sim 100$\,{\rm MeV}. Note that this is just the value 
preferred in Ref.~\refcite{ref26}. We emphasize that the authors 
of  Ref.~\refcite{ref26} just have taken into account power 
corrections. In standard analyzes of deep inelastic data 
(neglecting higher-twist effects) larger values of  $\Lambda$ are 
usually obtained. It is known, however, that if one includes in a 
phenomenological analysis the effects of higher twists, then it 
is possible\cite{ref37} to describe the data using an arbitrarily 
small  $\Lambda$.

Apart from power corrections related to the primordial transverse 
momentum of quarks (which correspond to operators involving the 
curl $G_{\mu\nu}$), there exist also power corrections due to 
twist-3 operators  $\psi \gamma_5 D^n \psi$ and $\psi \gamma_5 
\sigma_{\mu\nu}D^n \psi$.
In the large-$Q^2$  limit their 
contribution has an additional factor $\lambda^2/Q^2$  compared 
to the contribution of the twist-2 operators $\psi \gamma_5 
\gamma_\mu D^n \psi$. Note, however, that $\lambda$ is 
anomalously large 
\begin{align}
 \langle 0 | \bar d \gamma_5 u |P\rangle = i f_\pi
\frac{m_\pi^2}{m_u+m_d}\simeq i f_\pi\cdot (1.8\,{\rm GeV})\,, 
\label{eq4.9}
\end{align}
i.e. for $Q^2\lesssim 6\,{\rm GeV}^2$ these operators cannot be 
neglected\footnote{{\it (Comm. 2009)} B.~V.~Geshkenbein and M.~V.~Terentev, in 
{\it Phys.\ Lett.}   B {\bf 117},   243 (1982),   
showed that one should also include $\psi \gamma_5 \gamma_\mu D_\perp  D^n \psi$ 
operators.}. 
For the pseudoscalar $\psi\gamma_5 D^n\psi$-operator 
we have 
\begin{align}
 F_\pi^{PP} (Q)=\frac{4\pi\alpha_s(4M^2) }{N_c}\,C_F & 
\int\limits_0^1\varphi_P (x)\,\varphi_P(y)\di x\,\di y\, 
\nonumber \\[-5mm]
& \times\frac{1-x}{(xQ^2 +M^2)(xyQ^2+2 M^2)}\, 
 \left\{1+{\cal O}(\alpha_s)\right \}\,,  \label{eq4.10}
\end{align}
where $\varphi_P(x)=\lambda\varphi(x)\simeq\lambda f_\pi$.   

It should be noted that for $M^2=0$ the amplitude $E^{(PP)} 
(x,y)$ is as singular at $x \sim  0$  as $1/x^2$. As a result,  
integration over $x$ gives an additional factor $Q^2/M^2$  that 
compensates the absence of the $Q^2$-factor in the numerator of 
Eq. (\ref{eq4.10}). In  other words, in the region $Q^2\ll 
(M^2/m_q)^2$ the contribution  $F_\pi^{PP} (Q)$ has 
$1/Q^2$-behaviour rather than $1/Q^4$. Moreover, $F_\pi^{PP} (Q)$ 
has an additional large factor $(\lambda/M)^2 \gtrsim 10$ 
compared to  $F_\pi^{AA} $. The same factor has  also the 
$F_\pi^{TP}$-contribution ((TP) stands for $\psi\gamma_5 
\sigma_{\mu\nu}\psi\otimes\psi\gamma_5\psi$-projection): 
\begin{figure}[t]
\begin{center}
\includegraphics
[width=.75\textwidth]{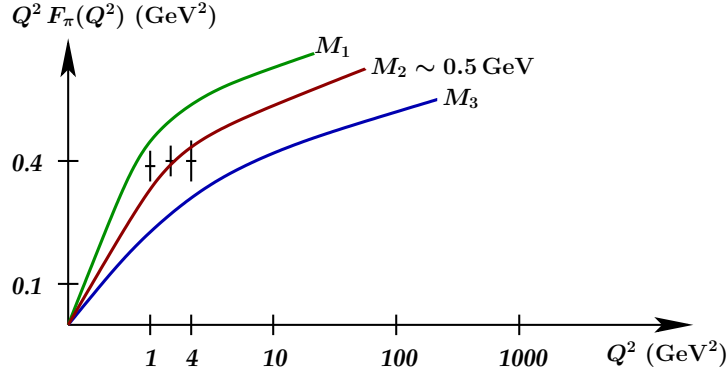}
\end{center}
\caption{\label{fig4}\footnotesize Dependence of pion form factor 
predictions on the confinement parameter $M$.}
\end{figure}
\begin{align}
 F_\pi^{TP}(Q)=& \frac{2\pi\alpha_s(4 M^2)\,C_F}{N_c}  
\int\limits_0^1\frac{\varphi_T(x)\,\varphi_P(y)\di x\,\di y} 
{(xQ^2+M^2)(xyQ^2+2M^2)}\, \nonumber \\
& \times\left(1-\frac{(1+x)Q^2}{xQ^2+M^2}-\frac{y(1+x)Q^2} 
{xyQ^2+2M^2}\right)\,\left\{1+{\cal O}(\alpha_s)\right\}\  .  
\label{eq4.11}
\end{align}
\begin{wrapfigure}[11]{R}{.5\textwidth}
\begin{center}
\vskip-5mm 
\includegraphics
[width=.45\textwidth]{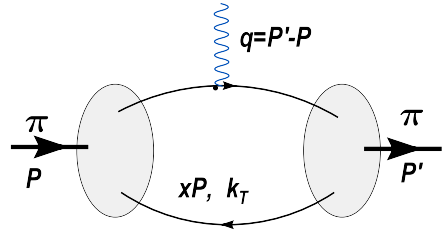}
\end{center}
\caption{\label{fig5}\footnotesize  Pion form factor  in light-front 
formalism.}
\end{wrapfigure}

Note that for $M^2=0$  the amplitude $E^{TP}$  has the 
$1/x^3$-singularity. However, using the equations of motion it is 
possible to show that the function $\varphi_T(x)$  has an 
additional $x$-factor for $x\sim 0$. In particular, if  
$\varphi_P(x) \sim \lambda f_\pi$, then \mbox{$\varphi_T(x) 
\sim\lambda f_\pi x(1-x)$}. As a result,  $F_\pi^{TP}(Q) \sim 
1/Q^2$ in the region $Q^2\lesssim(M^2/m_q)^2$. This contribution 
is negative and small for $Q^2\gtrsim4\,{\rm GeV}^2$. 
The curves for $F_\pi$ given by the sum of Eqs. (\ref{eq4.7}), 
(\ref{eq4.10}), (\ref{eq4.11}) have a right form (Fig. 
\ref{fig4}) and for $M^2\approx 0.1\div 0.2\,{\rm GeV}^2$, 
$\Lambda=100\,{\rm MeV}$ they are close to existing experimental 
data. 

It is easy to realize that since the passive quark in our case 
has a very small fraction of the pion momentum  (``wee'' parton), 
we deal really with the mechanism proposed by Feynman\cite{ref38} 
to explain the power-law fall-off of hadronic form factors. Thus, 
we might have considered the diagram shown in Fig. \ref{fig5} and 
write $F_\pi(Q)$ in the standard bound state 
formalism\cite{ref35,ref36}  
\begin{align}
 F_\pi(Q)\sim\int\limits_0^1\frac{\di x}{x (1-x)}\int\di^2k_\perp\,
\phi(1-x,k_\perp)\,\phi(1-x,k_\perp+x q_\perp)\,.\label{eq4.12}
\end{align}

Note that according to the Bethe-Salpeter equation the diagrams 
shown in Figs. 3 and 5 are equivalent up to ${\cal O}
(\alpha_s)$-corrections.  

\begin{figure}
\begin{center}
\includegraphics
[width=.75\textwidth]{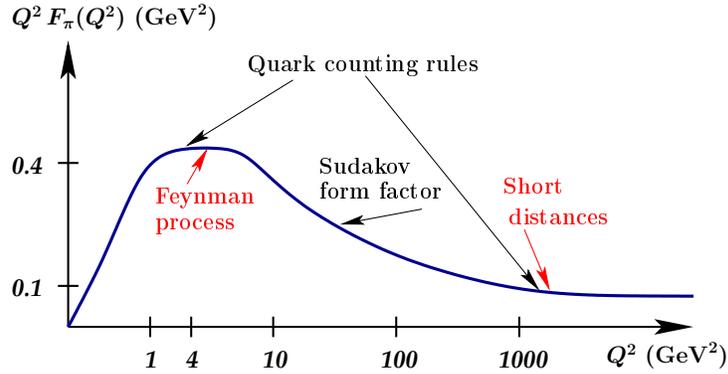}
\end{center}
\caption{\label{fig6}\footnotesize Prediction for pion
form factor with Sudakov form factor included.}
\end{figure}

If the function $\phi(x,k_\perp)$ is that given by Eq. 
(\ref{eq4.6}) then, performing $k_\perp$-integration in Eq. 
(\ref{eq4.12}) we obtain 
 \begin{align}
 F_\pi (Q)\sim\int\limits_0^1\di x\,\exp\left[-\frac{xQ^2}{2M^2(1-x)}
 -2\left(\frac{m_1^2}{xM^2}+\frac{m_2^2}{(1-x)M^2}\right)\right]\,,
\label{eq4.13}
\end{align}
whence it follows that for $Q^2<M^4/m_q^2$ the main 
contribution is given by the region $x\sim M^2/Q^2$. If the 
function $\varphi (x)$ (i.e. $\phi(x,k_\perp)$ integrated over 
$k_\perp$) behaves like $x^R$ for $x\sim 0$, then 
$F_\pi(Q)\sim(Q^2)^{-1-R}$. Our choice (\ref{eq4.5}) corresponds 
to $R\sim 0$, and as a result  $F_\pi(Q)\sim 1/Q^2$. If one 
assumes that 
 \begin{align}
  \chi_P (k_1,\ldots,k_n)\sim\exp\left\{\sum k_i^2/M^2\right\}
  \label{eq4.14}
 \end{align}
for a system composed by $n$ valence quarks, then $\varphi(x) 
\sim x^{n-2}$ for $x\sim 0$ and thus 
 \begin{align}
F_{(n)}(Q)\sim(Q^2)^{1-n}\,.\label{eq4.15}
 \end{align}
This relation corresponds formally to the well-known quark 
counting rule (QCR).\cite{ref39,ref40} In our case, however, this 
rule has nothing in common with short distances and scale 
invariance. The short-distance mechanism proposed in 
Ref.~\refcite{ref40} to explain QCR, according to our estimate, 
works only for $Q^2\gtrsim 10^3\,{\rm GeV}^2$. In an intermediate 
region one must take into account the fact that the contribution 
of the Feynman mechanism is damped by the Sudakov form factor of 
the active quark. Thus, one must multiply the curves shown in 
Fig. \ref{fig4} by the Sudakov QCD form factor\cite{ref41} 
 \begin{align}
  S(Q^2,M^2)=\exp\left\{-\frac{2C_F}{b_0}\left[
\left(\ln\frac{Q^2}{\Lambda^2}-\frac32\right)\left(\ln \frac{\ln 
Q^2/\Lambda^2}{\ln M^2/\Lambda^2}\right)-\ln\frac{Q^2}{M^2} 
\right ]\right\}\,.\label{eq4.16}
 \end{align}
For $\Lambda=100\,{\rm MeV},\ M^2=0.22\,{\rm GeV}^2$ this gives 
the curve shown in Fig. \ref{fig6}. In the region $Q^2=1 \div 
4\,{\rm GeV}^2$ there is good agreement with experimental 
data.\cite{ref42}  Decrease of  $Q^2 F_\pi (Q^2)$  for $Q^2\gtrsim 
10\,{\rm GeV}^2$ is due to the Sudakov form factor. In the region 
$Q^2\gtrsim 100\,{\rm GeV}^2$ the short-distance regime begins to 
work. In this region the average off-shellness of the gluon 
increases, $\mu^2$ grows, and as a result the wave function 
becomes narrower: 
 \begin{align}
  \varphi(x,\mu^2)\simeq\left(x\left(1-x\right)\right)^
{\frac{2C_F}{b_0}\left[\ln\ln\frac{\mu^2}{\Lambda^2}-\ln\ln  
\frac{\mu^2}{2M^2}\right] }\,, \label{eq4.17}
 \end{align}
and this, in turn, damps the contribution of the 
Feynman regime. For $Q^2\gtrsim10^3\,{\rm GeV}^2$ one may neglect 
the $F_\pi^{(PP)}$  contribution and use Eqs. (\ref{eq4.1}), 
(\ref{eq4.2}) with the wave functions $\varphi(x)\sim(x( 
1-x))^{0.2 \div 0.3}$. The asymptotic formula $\varphi(x)=6f_\pi 
x(1-x)$ may be used only for $Q^2 \gtrsim 10^{20}\,{\rm GeV}^2$. 

Thus, in the region $Q^2 \gtrsim 10^3\,{\rm GeV}^2$ begins the 
asymptopia, and  $F_\pi(Q)$ is again given by the quark counting 
rules. In this case they are due to the short-distance scale 
invariance, as expected in the pioneering works by Matveev, 
Muradyan, and Tavkhelidze\cite{ref39} and Brodsky and 
Farrar.\cite{ref40} 

Quark counting rules for ultra-relativistic systems were 
considered first by Terentyev.\cite{ref43}  However, he used 
constituent quark masses, $m_q\sim M$. In this case the range of 
applicability of our analysis ($M^2\ll Q^2\ll M^4/m_q^2$) is 
zero. We insist on using the current quark masses ($m_q\sim 4-7 \, 
{\rm MeV}$) in the wave function (\ref{eq4.6}). Note that the eventual 
IR cut-off in our analysis is of an order of $M$ (i.e. of an 
order of constituent quark mass) in accordance with the common 
wisdom. 

The last but not least observation is that the magnetic proton 
form factor in QCD is negative for narrow wave 
functions\cite{ref44} like $\varphi(x_1,x_2,x_3)\sim\prod
\delta(x_i -1/3)$ and positive for the broad ones, e.g. for that 
given by Eq. (\ref{eq4.14}). 

Summarizing this section, we may conclude that although our 
analysis is semi-phenomenological and some assumptions are very 
crude, it is clear, nevertheless, that a consistent treatment of 
power ``corrections''  (in fact, they give the main effect) is 
the main problem for perturbative QCD of hard elastic processes 
in the now (and, perhaps, forever) available energy range. 

\section{Soft processes and perturbative QCD} 
\label{sec5} 
\setcounter{equation}{0}

The main fraction of the total cross-section at high energies is 
due to the processes with small transverse momentum (soft 
processes). These are the elastic and quasi-elastic processes in 
the diffraction region $t\ll 1\,({\rm GeV}/c)^2$ and multiple 
production processes with low $p_T$: $p_T \ll 1\,{\rm GeV}/c$. The 
conventional phenomenology of processes in this region is the 
Regge-Mueller picture. Till now, however, it was not clear 
whether perturbative QCD can give any information about these 
processes. Below we discuss this problem analyzing a process 
${\bf 12\to1'2'}$ in the region $s\gg t,\,m^2_{\rm hadr}$. We 
will assume also that the $t$-channel is flavour nonsinglet. This 
assumption greatly simplifies the analysis. 

For scalar gluons, i.e. in a Yukawa-type field theory soft 
processes have been studied 10 years ago.\cite{ref45}  It was 
shown that summation of all logarithmic terms $(\log S)^N$ coming 
from the short-distance integration (regime 1, see Section 
\ref{sec2}) gives the following representation 
\be
f^{\pm}(j,t)=C^{\pm}(j,t)[l-B^{\pm}(j,t)]^{-1}v^{\pm}(j)C^{\pm}(j,t) 
\label{eq5.1}
\ee
for the Mellin transform of the scattering amplitude 
$F^{\pm}(S,t)$:
\be
F^{\pm}(S,t)=\frac{1}{2i}\int\limits_{-i\infty}^{i\infty}\!\!\!\di j 
\,\frac{|S|^j(e^{i\pi j}\pm 1)}{\Gamma(j+1)\sin(\pi j)}f^\pm(j,t)  \  , 
\label{eq5.2} 
\ee
where $\pm$ stands for signature, $C,\,v\mbox{ and } B$ are some 
matrices (to be discussed below); e.g., $B=B_{ab}$, and 
$a,b=S,V,T,A,P$ are structures appearing in the Fierz identity 
applied to factorize the spinor structure of the relevant 
contributions. According to this representation (Eq. 
(\ref{eq5.1})) the Mellin transform $f^\pm(j,t)$ possesses moving 
($t$-dependent) Regge poles due to zeros of ${\rm Det} 
[1-B(j,t)v(j)]$. It has also fixed ($t$-independent) 
singularities in the complex $j$-plane accumulated in the 
function $v(j)$. The type of fixed singularities depends on the 
ultraviolet asymptotics of the effective coupling constant. In 
particular, in a fixed point theory (where $\bar g(\mu)\to g_0$ 
as $\mu\to\infty$) the function $v(j)$ has square-root branch 
points, the position of which depends on $g_0$,  i.e. on the 
asymptotical value of $\bar g(\mu)$. On the other hand, in an 
asymptotically free field theory $v(j)$ has the infinite number 
of poles condensing to $j=0$. 

Let us now discuss briefly the derivation of Eq. (5.1). 
Consider a particular diagram of a binary process 
${\bf 12 \to 1'2'}$ in the region 
$s\gg|t|,\,m_{\rm hadr}$ The Mellin transform 
of its contribution has the following structure in the 
$\alpha$-representation (Eq. (\ref{eq3.2})): 
 \be 
f^\pm(j,t)\propto \int\prod\limits_\sigma\di\alpha_\sigma 
D^{-2}(\alpha)g(j,t,\alpha)|A(\alpha)|^j [\theta(A)\pm\theta(-A)]
\exp[i J(\alpha,t,m^2)]\,,
 \label{eq5.3} 
 \ee
where $g(j,t,\alpha)$ is a polynomial in $j$ (it corresponds to 
the function $G$ in Eq. (\ref{eq3.2})) and $A$ is the coefficient 
in front of the large variable $S=s-u$. As is well known, the 
asymptotical behaviour of $F(S,t)$ for large $S$ is determined by 
the rightmost singularities of its Mellin transform $f(j,t)$. 
These are poles $j-N$ generated by integrations corresponding to 
the regimes 1)--3) discussed in Section \ref{sec3}. However, 
using Eq. (\ref{eq3.3}) it can be shown that in Yukawa theory the 
IR-regime gives only non-leading poles at $j=-1,-2,\dots$. 
Furthermore, the pinch regime contributes only to the negative 
signature amplitude $F^-(S,t)$. Thus, for $F^+(S,t)$ it is 
sufficient to consider only the poles due to the short-distance 
regime. 

\begin{figure}[t]
\begin{center}
\includegraphics[width=.9\textwidth]
{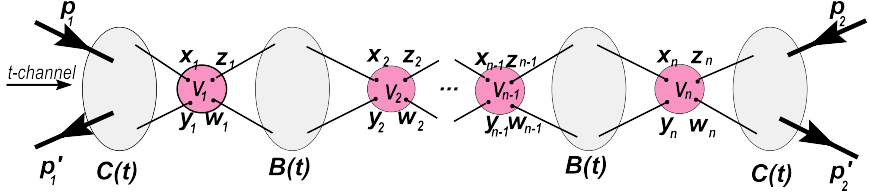}
\end{center}
\caption{\label{fig7}\footnotesize 
Illustration of Equation (\ref{eq5.4}).}
\end{figure}

\begin{figure}[b]
\begin{center}
\includegraphics[width=.9\textwidth]
{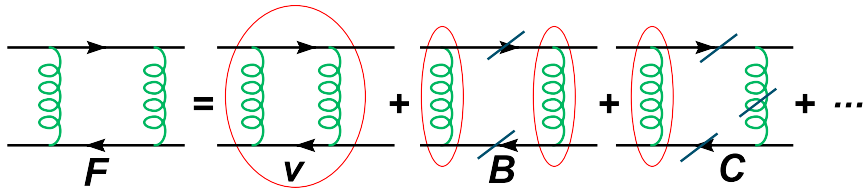}
\end{center}
\caption{\label{fig8}\footnotesize 
The leading terms of a  particular box diagram.}
\end{figure}

According to Eq. (\ref{eq3.3}), the leading poles (at $j=0$) are 
due to the subgraphs $V_i$ with 4 external lines. We recall that 
$V_i$ should possess the property that if it is contracted into 
point, the diagram becomes $S$-independent (i.e. $V_i$ must be an 
\mbox{$S$-subgraph)}. The most general configuration is shown in 
Fig. \ref{fig7}. Note that in general the SD-subgraphs $V_i$ may 
be 2-particle-reducible, i.e. they may contain smaller 
$S$-subgraphs with 4 external lines and the total singularity due 
to the SD-regime of $V_i$ may be a multiple pole $j^{-N_i}$. It 
makes sense to treat a particular diagram as a ladder composed by 
2-particle irreducible blocks $k_j$. Then the maximal value of 
$N_i$ is determined by the number of $k_j$'s inside $V_i$ (and 
also by the number of the UV-divergent subgraphs inside $V_i$). 
The contribution $f_V(j)$ of each $S$-subgraph $V$ may be 
represented as a sum of two terms $f_V=f_V^{\rm pole}+f_V^{\rm 
reg}$. The first term ($f_V^{\rm pole}$) is due to integration 
over the region $\sum\alpha_\sigma\equiv\lambda_V<1/\mu^2$ and 
the second one is due to that over the region 
$\lambda_V>1/\mu^2$. This procedure corresponds to a subtraction 
of the pole due to the small $\lambda_V$ integration. However, if 
$V$ is composed by two or more $k_J$'s then $f^{\rm reg}$ may 
also possess the poles at $j=0$ due to the SD-integration for a 
smaller subgraph $V_i\subset V$. Thus, one must represent 
$f_V^{\rm reg}$   as $f_V^{\rm reg}V_i^{\rm pole}+f_V^{\rm 
reg}V_i^{\rm reg}$, and so on. An example of such a decomposition 
is shown in Fig. \ref{fig8}, where the pole parts are circled by 
the thin (red) line and the regular ones marked by the slashed 
(blue) lines. Note that Fig. \ref{fig7} is really a decomposition 
of the whole diagram. Summing over all diagrams we obtain (in the 
coordinate representation):
\ba
\label{eq5.4}
(2\pi)^4\delta^4(p_1+p_2-p'_1-p_2')F(S,t)&=&
\sum\limits_{n=1}^\infty\int C(p_1,p'_1;x_1,y_1)\\[-3mm]
\prod\limits_{i=1}^{n-1}\left\{[\di\chi_i]\,v(x_i,y_i;z_i,w_i)
B(z_i,w_i;x_{i+1},y_{i+1})\right\}
&&\!\!\!\!\!\!\!\!v(x_n,y_n;z_n,w_n)C(p_1,p'_1;z_n,w_n)[\di\chi_n]
\nonumber
\ea
where $[\di\chi]_i\equiv\di x_i\di y_i\di z_i\di w_i$ (see Fig. 
\ref{fig7}), and the functions $C,\,v,\, B$ are given by the 
following matrix elements 
\ba
C_a(p_1,p'_1;x_1,y_1)&=&R_{\mu^2}\la p'_1|S^+T(:\bar\psi(x_1)\Gamma_a
\psi(y_1):S)|p'_1\ra\,,\label{eq5.5}\\
\hspace{-8mm} B_{ab}(z_i,w_i;x_{i+1},y_{i+1})\!\!&=&\!\!R_{\mu^2}\la0|S^+T(:\!\!
\bar\psi(z_i)\Gamma_a\psi(w_i)\!\!:\ :\!\!\bar\psi(x_{i+1})\Gamma_b
\psi(y_{i+1})\!\!:S)|0\ra\,,\label{eq5.6} \\
v(x_i,y_i;z_{i+1},w_{i+1})\!\!&=&\!\!P_{\mu^2}\la0|S^+T(:\!\!
\bar\eta(x_i)\Gamma_a\eta(y_i)\!\!:\ :\!\!\bar\eta(z_{i+1})\Gamma_b
\eta(w_{i+1})\!\!:S)|0\ra\,,\label{eq5.7}\\
\Gamma_a,\,\Gamma_b&=&1,\,\gamma_\mu,\,\sigma_{\mu\nu},\,\gamma_5
,\,\gamma_5\gamma_\mu\,. \label{eq5.8}
\ea 
Here, $:\ \ :$ denotes the usual normal product; 
$\eta,\,\bar\eta$ are the spinor currents (e.g. \mbox{$\eta=S^+(\delta 
S/\delta\psi))$;} $S$ is the S-matrix; $P_{\mu^2}$ means that 
$\lambda_V< 1/\mu^2$ for each diagram $V$ contributing to $v$, 
and $R_{\mu^2}$ means that $\lambda_V > 1/\mu^2$ for all leading 
$S$-subgraphs having lines related to B or C. 

If we expand $B$ or $C$ into the Taylor series over $\xi=x-y$, 
$\zeta=z-w$, 
\ba
&&C_a(p_1,p'_1;x_1,y_1)\!=\!\exp(irX_1)R_{\mu^2}\!\!\!
\sum\limits_{j=0}^\infty
\frac1{j!} \la p'_1|S^+T(O_{a\mu_1\cdots\mu_j}S)|p'_1\ra
\xi_1^{\mu_1}\!\!\cdots\xi_1^{\mu_j},\label{eq5.9}\\ 
&&B_{ab}(z_i,w_i;x_{i+1},y_{i+1})= \label{eq5.10}\\ \nonumber
&&R_{\mu^2}\!\!\sum\limits_{j=0}^\infty\!\sum\limits_{k=0}^\infty
\frac1{j!k!}\, \xi_i^{\nu_1}\!\!\cdots\xi_i^{\nu_j}\zeta_{i+1}^{\mu_1}\!\!
\cdots\zeta_{i+1}^{\mu_k}\la0|S^+T[O_{a\nu_1\cdots\nu_j}(Z_i)
O_{b\mu_1\cdots\mu_k}(X_{i+1})S]|0\ra,
\ea
(where $r=p_1-p_1'$, $X_i=\frac{x_i+y_i}{2}$, 
$Z_i=\frac{z_i+w_i}{2}$), then $R_{\mu^2}$ provides the 
renormalization recipe for the resulting composite operators 
$O_{a\nu_1\cdots\nu_j}=\bar\psi\Gamma_a\partial_{\nu_1} \cdots 
\partial_{\nu_j}\psi$. It should be emphasized, however, that 
$R_{\mu^2}$ in addition, subtracts from $B,\,C$ also the 
contributions due to integration over small 
$\lambda_V$-parameters for $S$-subgraphs that do not contain the 
vertices $\bar\psi\Gamma\partial^n\psi$ corresponding to 
composite operators. 

Just like for hard processes, only the lowest-twist operators (i 
e. the traceless-symmetric part $O_{\{a\nu_1\cdots\nu_j\}}$ of 
$O$)  give the leading contributions. Using the translation 
invariance of the functions (\ref{eq5.5})--(\ref{eq5.7}),
integrating over $\xi_i,\,\zeta_i,\,X_i$ and $Z_i$ and summing 
over $n$ ($n$ is the number of SD-integrations), we obtain 
\be
F(S,t)=\sum\limits_{j=0}^\infty\frac{|S|^j}{j!}C(j,t)\cdot
\left[1-v(j)B(j,t)\right]^{-1}v(j)\cdot C(j,t)
\label{eq5.11}
\ee
where $t=r^2$, $S=2(PQ)$, $P=p_1+p_1'$, $Q=p_2+p_2'$. The functions 
$B(j,t),\,C(j,t)$ and $v(j)$ are given by 
\ba
&&R_{\mu^2}\la p'_1|S^+T(O_{a\mu_1\cdots\mu_j}S)|p'_1\ra=
C(j,t)\{P_aP_{\mu_1}\cdots P_{\mu_j}\}+{\cal O}(r_\mu)\label{eq5.12}\\
&&\int\di X e^{irX}R_{\mu^2}\la0|S^+T[O_{a\nu_1\cdots\nu_j}(Z_i)
O^{b\mu_1\cdots\mu_k}(X_{i+1})S]|0\ra=  \nonumber\\
&&\hskip55mm j!\delta_k^j\delta_{\{a}^{\{b}\delta_{\mu_1}^{\nu_1}\cdots
\delta_{\mu_k\}}^{\nu_j\}}B(j,t)+{\cal O}(r_\nu)\label{eq5.13}\\
&&\int\di(X-Z)\di\xi\di\zeta\exp[ir(X-Z)]_{r^2=0}\tilde{v}(X-Z,\xi,\zeta)
\xi^{\nu_1}\cdots\xi^{\nu_j}\zeta_{\mu_1}\cdots\zeta_{\mu_k}= \nonumber\\
&&\hskip60mm j!v(j)\delta_k^j\delta_{\{\mu_1}^{\{\nu_1}\cdots
\delta_{\mu_k\}}^{\nu_j\}}+{\cal O}(r_\nu) \  , \label{eq5.14}
\ea 
where $\tilde{v}(X-Z,\xi,\zeta)$ is defined by
\be
v(x,y,z,w)= v \left (X+\frac{\xi}{2},X-\frac{\xi}{2},Z+\frac{\zeta}{2},
Z-\frac{\zeta}{2} \right )=\tilde{v}(X-Z,\xi,\zeta)
\label{eq5.15}
\ee
and ${\cal O}(r_\nu)$  denotes terms containing $r_\nu$. These 
give zero contribution into Eq. (\ref{eq5.11}) because 
$(Pr)=(Qr)=0$. Note that Eqs. (\ref{eq5.1}), (\ref{eq5.2}) give 
just a Mellin transformed version of Eq. (\ref{eq5.11}). 

To construct the functions $B,\,C$ and $v$ one must apply first 
the R-operation for ordinary divergent subgraphs (this procedure 
is characterized by the renormalization parameter $\mu_R$) and 
then the operations $P_{\mu^2}$ and $R_{\mu^2}=1-P_{\mu^2}$ that 
separate small and large $\lambda$-parameters (this corresponds 
to splitting of the mass logarithms $\ln S/p^2$ into 
``short-distance'' $(\ln S/\mu^2)$ and ``long-distance''  
$(\ln\mu^2/p^2)$ parts). The whole amplitude $F(S,t)$, of course, 
must be independent both of $\mu$ and $\mu_R$. The 
$\mu_R$-independence of $F$ leads to a standard renormalization 
group equation 
\be
(\mu_R\partial/\partial\mu_R+\beta(g)\partial/\partial g-
4\gamma_\psi)v(j,\mu_R,g,\mu)=0  \  .
\label{eq5.16}
\ee 

The subtraction procedure $R_{\mu^2}$ for the problem considered 
is more complicated than that for hard processes. This is mainly 
due to the fact that, for soft processes, we deal in general with 
the configuration (Fig. \ref{fig7}) that has several 
non-overlapping SD-subgraphs $V_1,\dots V_ n$. We recall that for 
hard processes we always have a configuration with the only 
SD-subgraph (see, e.g., Fig.\ref{fig2}a). Straightforward 
analysis gives the following equation for C 
\be 
\mu\frac{\di}{\di\mu}C\equiv C'=\gamma(j)C+(1-B v)^{-l}(1-B v)'\,,
\label{eq5.17}
\ee 
where all functions entering into Eq. (\ref{eq5.17}): $B,\,C$ and 
$v$ depend on $\mu$. The second term in Eq. (\ref{eq5.17}) is 
just due to the additional subtraction discussed above. The 
function $\gamma(j)$ is the ordinary anomalous dimension of the 
composite operator $O_{a\nu_1\cdots\nu_j}$. In our case it is 
convenient to single out from $\gamma$ the terms singular at 
$j=0$. It can be shown that these terms are proportional to $v$, 
i.e. that $\gamma(j)=c(j)+b(j)v(j)$,  where $c(j)$ is regular at 
$j=0$.

The equation similar to Eq. (\ref{eq5.17}) can be obtained also 
for $(1 - Bv)$: 
\be 
-(1-Bv)'= [2\gamma(j)+bv(1-Bv)+Bv'][1-Bv]. 
\label{eq5.18}
\ee
 
Requiring that $\Phi^{\rm pole}(j)$, the sum of the leading 
poles, does not depend on $\mu$ we obtain that $\Phi^{\rm 
pole}(j)$, must be regular at $j = 0$: 
\be
\mu\frac{\di}{\di\mu}\Phi^{\rm pole}(j)=r(j)\,,  
\label{eq5.19}
\ee
where $r(j)$ is some function regular at $j=0$. Using Eqs. 
(\ref{eq5.17}-\ref{eq5.19}) we obtain the equation for $v$:
\be 
v'+2\gamma v+bv^2=-r\,.
\label{eq5.20}
\ee 
The meaning of $r(j)$ becomes clear, it is just the residue of 
$v(j)$ at $j=0$, because $v'\equiv jv$. 

It should be remarked that Eq. (\ref{eq5.20}) differs from its 
analogue given in Ref.~\refcite{ref45} because of another choice of 
the $R_\mu^2$-operation.   

Using Eq. (\ref{eq5.20}) one can sum up all the poles at $j=0$ 
due to the SD-regime of all possible $S$-subgraphs (i.e. to sum 
all $\log^N(S/p^2)$ contribution). The solution of Eq. 
(\ref{eq5.19}) has square root branch points in the complex 
$j$-plane\cite{ref45} (see also Ref.~\refcite{ref46}). However, 
$v(j)$ has also  poles due to divergent subgraphs. These poles 
(i.e. $\log^N(S/\mu_R^2)$-contributions) are summed by Eq. 
(\ref{eq5.16}). If we take $\mu=\mu_R$ and combine Eqs. 
(\ref{eq5.16}), (\ref{eq5.20}), we obtain  
\be
\beta(g)\frac{\partial v}{\partial g}=(j-2\gamma-4\gamma_\psi)v- 
b v^2-r\,,
\label{eq5.21}
\ee 
where $v,\,\gamma$ and $r$ depend on $j$, and $(-j)$ is the 
canonical dimension of $v$. In the lowest order of PT $b=1,\, 
r\sim\gamma \sim\gamma_\psi\sim g^2$, $\beta(g)\sim g^3$, and the 
solution of Eq. (\ref{eq5.21}) has condensing poles at $j=0$.\cite{ref45,ref46}

Summarizing the preceding discussion, we conclude that if one 
assumes that the asymptotical behaviour of $F(S,t)$ is given by 
the sum of the leading terms of all contributing Feynman 
diagrams, then $F(S, t)$ has for large $S$ a Regge-type behaviour 
$F(S,t)\sim C^2(t)S^{\alpha(t)}$ since its Mellin transform has 
just a $t$-dependent singularity at $j=\alpha(t)$. To find the 
function $\alpha(t)$ explicitly, one must solve the equation
\be 
{\rm Det}[1 -B(j,t,\mu_R,g,\mu,m)v(j,\mu_R,g,\mu)]=0\,.
\label{eq5.22}
\ee 
It can be shown that Eq. (\ref{eq5.18}) guarantees that 
$\alpha(t)$ does not depend on $\mu$ and $\mu_R$:
\be 
\alpha(t)= \phi(m_q^2/t,t/\mu^2,\bar{g}(\mu^2))=\phi(m_q^2/t,1,\bar{g}(t))
\label{eq5.23}
\ee 

Hence, one may try to calculate the Regge trajectories in the 
region where $g(t)$ is small, e.g. in QED, where 
$\alpha_{\mbox{\tiny QED}}\approx1/137$, or in QCD for 
sufficiently large $t$. However, there arises a question whether 
Eq. (\ref{eq5.11}) is valid in vector theories. 

In QCD one encounters the complication discussed in Section 
\ref{sec3}. First, a SD-subgraph $V_i$ may have an arbitrary 
number of external gluon lines. But if the \mbox{$t$-channel} is 
colour singlet, then the only change is
\be
\bar\psi(x)\Gamma\psi(y)\to \bar\psi(x)\Gamma P\exp\left(ig\int\limits_y^x
\hat A_\mu(z)\di z^\mu(y)\right)\psi(y)   
\label{eq5.241}
\ee  

\begin{wrapfigure}[11]{R}{.5\textwidth}
\begin{center}
\vskip-27mm 
\hspace{-15mm} 
\includegraphics[width=.5\textwidth]
{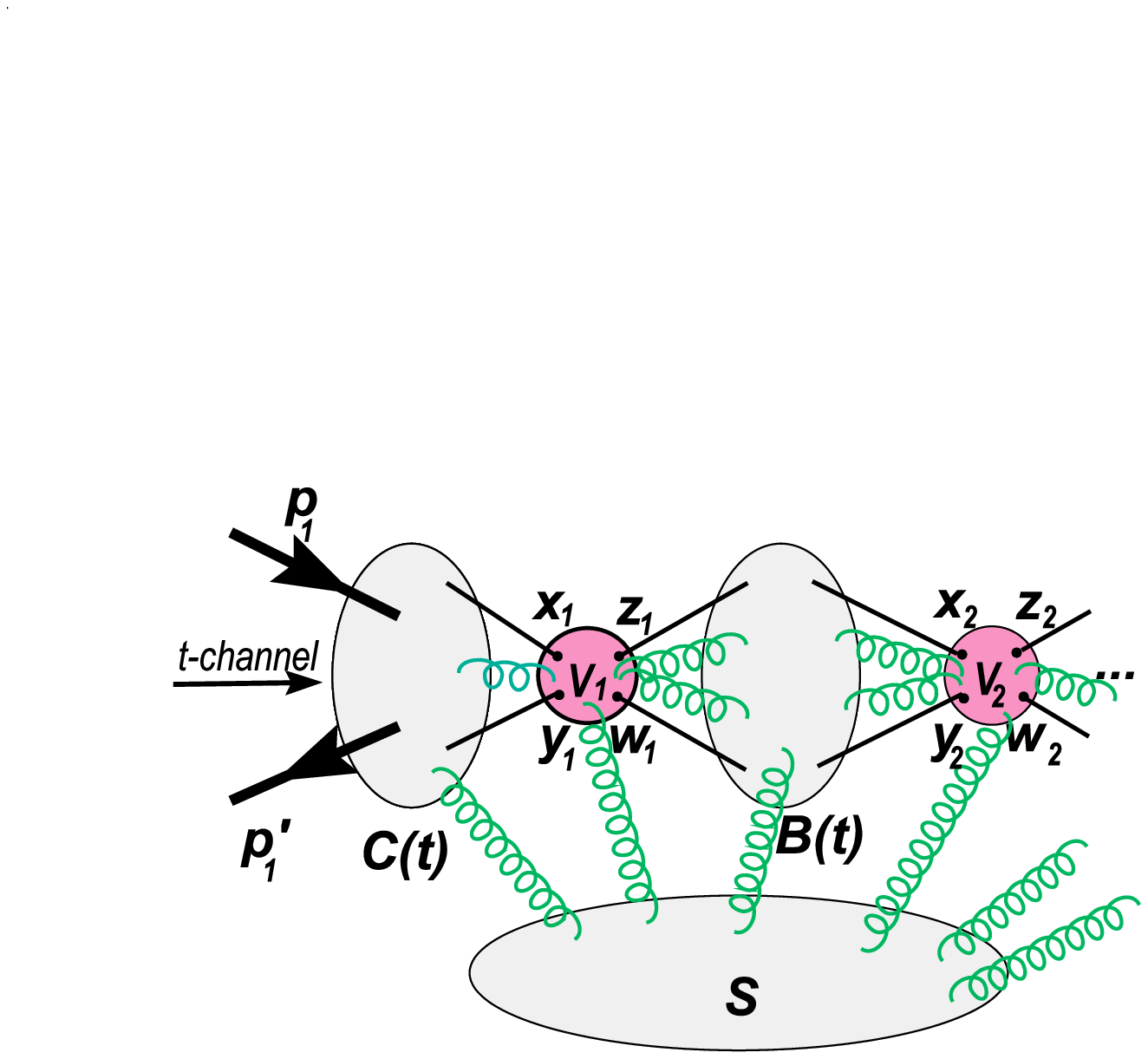}
\end{center}
\caption{\label{fig9}\footnotesize 
The same as Fig. \ref{fig7} but for the case of QCD.}
\end{wrapfigure}
\noindent 
for all bilocal operators entering into $B$- and 
$C$-functions. For local operators this corresponds to the change 
$\partial_\mu\to D_\mu=\partial_\mu-i g\hat A_\mu$. The second 
complication is due to the IR-regime (soft exchanges, see Fig. 
\ref{fig9}). However, just like for hard processes, if the 
$t$-channel is colour singlet, then the sum of all soft exchanges 
give only power corrections in each order of perturbation theory. 
Thus, all terms responsible for the leading power contribution 
have the structure of Fig. \ref{fig7} and as a result, we get 
Eq.~(\ref{eq5.11}). In other words, if we sum the leading $j$ 
singularities of all relevant Feynman diagrams in QCD, we obtain 
a Regge-type picture for binary processes and, hence, a 
multiperipheral picture for the multiple production at low $p_T$. 

We recall, however, that {\it we have discussed above only the 
flavour nonsinglet, positive signature amplitude $F^+_{NS}$}. For 
$F^-_{NS}$ the pinch regime (see Section \ref{sec3}) also gives 
leading $j$-poles for non-planar diagrams. It is known, however, 
that the non-planar diagrams have an additional colour factor 
$(1/N_c)^2=(1/3)^2$. This suggests that the pinch contributions 
in QCD must be suppressed. There exists also an experimental 
evidence in favour of this suppression: the well-known signature 
degeneracy of the Regge trajectories. 

For flavour singlet amplitudes $F_S$ (``vacuum'' exchange) the 
poles generated by the pinch regime are at $j=1$ rather than at 
$j =0$ due to the 2-gluon intermediate states and after summation 
one obtains for $F^+_S$ a square-root branch point\footnote{({\it 
Comm. 2009.}) The asymptotical freedom can, of course, change 
this singularity to a sort of condensed poles near $j\approx1$.} 
at \mbox{$j=1+{\cal O}(g^2)$.} This suggests that the pinch 
regime plays a highly important role in formation of the Pomeron 
singularity. 

The most intriguing possibility is to utilize the asymptotic 
freedom of QCD for a calculation of the Regge trajectories and of 
the resonance masses in the region of large $t$ (see Eq. 
(\ref{eq5.23})). Note, however, that the function $B(t)$ 
describes the long-distance dynamics, i.e. by its construction, 
$B$ has an UV cut-off but there is no IR cut-off. This means that 
if the IR region of integration gives a sizeable contribution, 
one must (in some way) take into account nonperturbative effects. 
It seems that the most effective tool here is the method proposed 
in Ref.~\refcite{ref26}. This and related problems are under 
investigation now.\footnote{({\it Comm. 2009.}) These  problems are 
still waiting for their investigators!}


\end{document}